# First Principle Study of Alkaline Metal Intercalation in Twisted Bilayer Graphene

N K R Perera, K M Abeywickrama, W W P De Silva*

*Department of Physics, University of Sri Jayewardenepura*
*wasanthidesilva@sjp.ac.lk

## ABSTRACT

The intercalant alkaline metals, Ca, Sr, & Ba, are sources of a periodic perturbation on the bilayer graphene, which leads to the formation of N-type semiconductor. The twist angle between graphene layers applies disorder to the intercalated bilayer graphene to tune the bandgap. The twisted bilayer graphene structures of $C_8MC_8$ (M: Ca, Sr, Ba) were studied for the Energy band diagrams, Density of States, transport properties of Seebeck coefficient, Hall coefficient, Thermal conductivity, and Electrical conductivity using Quantum-Espresso and BoltzTrap software. The bandgap enhances with respect to an increase in twist angle to values of technological requirements to implement in real applications. The peak of the Seebeck coefficient & Hall Effect increase, and Thermal Conductivity & Electrical Conductivity decrease for different twist angles. Therefore, $C_8MC_8$ has a significant impact of the twist angle, hence gives possible tailoring of material properties.

## 1. INTRODUCTION

The specific electronic structure of monolayer graphene has been well known for zero-gap massless electrons[1], and high carrier mobility[2], which leads to a limitation in real-world applications. Bilayer graphene (BLG) with different doping materials and applying electrical fields[3] was introduced to modulate the bandgap of graphene from zero to a few eV, which creates a high on-off ratio with a promising structures enabling to fabricate the fundamental electronic device of transistors. A single millimeter square sheet of BLG can be used to fabricate millions of differently tuned electronic devices due to tunable bandgap. The laser-induced BLG acts as a conductor or semiconductor, which trends to invent the computer chips made of 2D materials[4].

The graphene-metal intercalation is one of the other promising systems that can integrate with many practical applications. The theoretical study of electrical, thermal, and optical properties of intercalated graphene gives an insight into the device fabrications and new synthesis processes. The charge transfer to the derived π state of graphene, hybridization of the valence bands, and lattice mismatch between graphene-metal interfaces are the factors that influence changes in the electronic properties of graphene-metal systems. The metal atoms or molecular layers are inserted between graphene layers in BLG, which is called intercalation. Li- intercalated BLG has been experimentally investigated for crystallographic and electronic properties, and this material opens the gate to nano-scale ion battery[5]. The in-situ electrical transport measurement of Ca-intercalated BLG ($C_6Ca$) showed an interesting property of superconductivity at a critical temperature of 11.5 K[6]. Due to this Ca intercalation occurrence of two-dimensional $\pi$ bands of the Interlayer state (IL)[7] were understood as the hybridized state of metal's s-orbital and free-electron state. The presence of IL state indicator is responsible of having superconductivity in Bulk graphene layers[7]. However, there is no superconductivity property in Li-intercalated graphene.

Twistronic is another highlighted concept introduced by S. Carr et al.[8], to enhance different properties in 2D materials, including graphene bilayers and multilayers. The study of twistronic is experimentally significant when the fabrications of bilayers, usually there is a misalignment in rotation angle with the adjacent layers, which shows Moire patterns in Scanning Tunneling Microscope (STM) measurements. However, the twist bilayer structure has a complex geometry consisting of AA stacked and AB stacked locally. Due to the small twist angles, the flat bands responsible for forming a strongly correlated electron system, which leads to the remarkable property of superconductivity at the magic angle 1.1, was experimentally discovered in 2018[9]. When the twist angle is greater than the magic angle, there is a possibility of the opening bandgap in addition to the flat band formation, including an interesting electrical and optical properties. The graphene bilayers with a large twisting angle lead to decoupling in inter-van der Waals interaction that survives quasi-particle dynamics of the twisted bilayer as of single-layer graphene.

This study implements the concepts of intercalation and twisting in bilayer graphene. The alkaline metals of Ca, Sr, and Ba are considered. The band diagrams were studied based on the first principle calculation using Quantum espresso software[10], and based on that results, transport properties are calculated using Boltztrap software[11]. The electronic transport parameters of the Seebeck coefficient, Hall coefficient, Electrical conductivity and Thermal conductivity behaviors were studied to investigate the thermoelectric materials. These parameters are based on the charge mobility and phonon interaction of the material.

## 2. COMPUTATIONAL METHOD

### 2.1 Density Functional Theory (DFT)

DFT is a widely used computational model to deal with complex many-body problems using Kohn-Sham (K-S) (Eq. 1). The many-body Hamiltonian follows the first and second Hartree-Fock approximations of the mean-field approach to deduce the K-S equation. The K-S addresses the complicated e-e interactions with the same density of electrons ($n(r)$) in the system and models the system as a non-interacting system, as shown in equation 1.

$$\left(\frac{\hbar^2}{2m}\nabla^2 + v_{eff}(r)\right)\varphi(r) = \varepsilon\varphi(r) \ldots\ldots\ldots\ldots (1)$$

$$v_{eff}(r) = V_{ion}(r) + V_{xc}(r) + V_H(r) \ldots\ldots\ldots\ldots\ldots (2)$$

$$n(r) = N \int d^3r_1 \int d^3r_2 \ldots \int d^3r_N \Psi^*(r,r_1,r_2,\ldots,r_N)\Psi(r,r_2,\ldots,r_N) \ldots (3)$$

First Principle Study of Alkaline Metal Intercalation in Twisted Bilayer Graphene

Where, $v_{eff}$ is potential due to *e-iron* interactions, $V_{ion}(r)$ is the potential due to ironic cores, and $V_H$ is the potential due to other electrons distribution and $V_{xc}$ is the exchange-correlation potential, which has different approximations in the different calculations to improve the accuracy of the results. As shown in equation 3, the electron density is considered a fundamental variable in the DFT approach instead of individual particle wave function to find the materials' ground state properties.

### 2.2 BoltzTrap

The semi-classical Boltzmann theory has been incorporated in BoltzTrap software. The BoltzTrap approach considered the Fourier expansions of the band energies and obtained the group velocity, inverse mass tensor, as the first and second derivative of the bands with respect to momentum space *k*. The semi-classical Boltzmann equations and the transport tensors can be calculated based on the following equations.

$$\sigma_{\alpha\beta}(\varepsilon) = \frac{e^2}{N} \sum_{i,k} \tau_{i,\vec{k}} \vartheta_\alpha(i,\vec{k}) \vartheta_\beta(i,\vec{k}) \delta(\varepsilon - \varepsilon_{i,k}) \ldots\ldots\ldots\ldots (4)$$

$$S_{\alpha\beta}(T,\mu) = \frac{1}{eT\Omega\sigma_{\alpha\beta}(T,\mu)} \int \sigma_{\alpha\beta}(\varepsilon)(\varepsilon - \mu) \left[ -\frac{\partial f_0(T,\varepsilon,\mu)}{\partial \varepsilon} \right] d\varepsilon \ldots\ldots (5)$$

$$\sigma_{\alpha\beta}(T,\mu) = \frac{1}{\Omega} \int \sigma_{\alpha\beta}(\varepsilon) \left[ -\frac{\partial f_0(T,\varepsilon,\mu)}{\partial \varepsilon} \right] d\varepsilon \ldots\ldots\ldots\ldots (6)$$

Where $e$ is a charge of an electron, $\tau$ is the relaxation time which is constant in this code, N is the number of k-points, $\alpha, \beta$ are tensor indices, $v_\alpha, v_\beta$ are the group velocities, $\Omega$ is volume of the unit cell, $f_0$ is Fermi-Dirac distribution function. The equation 5 and 6 can be used to evaluate the Thermal conductivity and Electrical conductivity as a function of temperature and chemical potential, respectively.

## 3. RESULTS AND DISCUSSION

### 3.1 Alkaline Materials Intercalation in Bilayer Graphene

The computationally possible smallest system of $C_8MC_8$ was considered in this calculation, as shown in figure 1 (a). M is intercalated Alkaline metal atoms of Ca, Sr, and Ba placed at the middle of the carbon ring. The side view of $C_8MC_8$ is shown in figure 1(b). The placement of metal atoms selected as the middle position in the bilayer graphene is favorable for the lowest energy of the system. A unit cell of $C_8MC_8$ can be expanded to obtain a supercell structure of metal intercalated bilayer graphene, as shown in figure 1 (c).

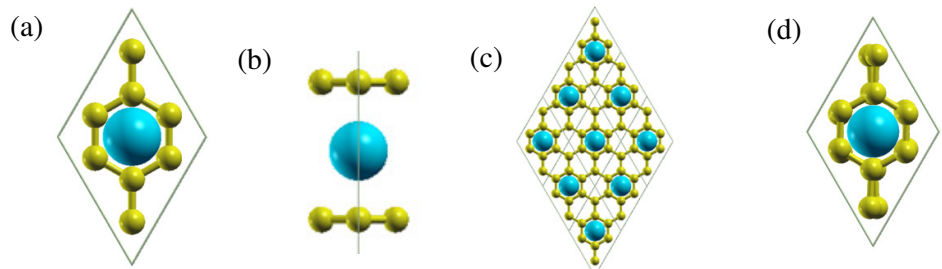

**Figure 1:** (a) $C_8MC_8$ structure (b) Side view of $C_8MC_8$ (c) Top view of metal intercalated bilayer graphene super-cell (d) Twisted $C_8MC_8$

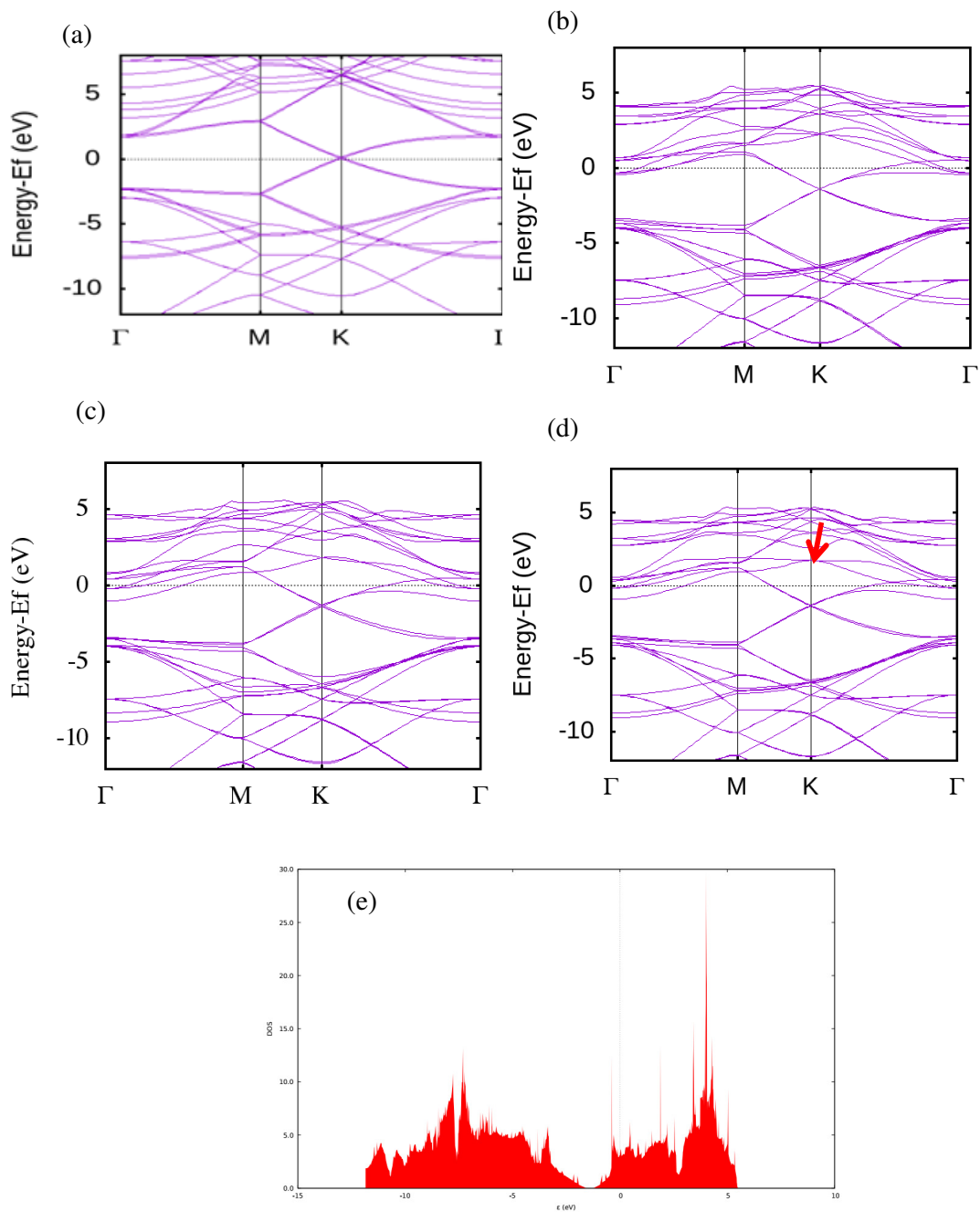

**Figure 2:** The band diagrams for (a) Bilayer graphene (b) $C_8CaC_8$ (c) $C_8SrC_8$ (d) $C_8BaC_8$ (e) Dos for $C_8CaC_8$.

The band structure calculation is essential to evaluate the effective mass and the transport properties of the material. The band structure diagrams for bilayer graphene, $C_8CaC_8$, $C_8SrC_8$, and $C_8BaC_8$ are shown in figure 2. Compared to the bilayer graphene band structure, there are some changes in the electronic structure, and its charge carrier densities due to Alkaline metal intercalation. As an example, Ca donate charges to the graphene surface. The donating nature is increased by Ca, Ba, Sr in this order. The $\pi$ and $\pi^*$ are shifted downward from the Fermi level at the $K$-point due to the electrons donation from Alkaline metal atoms to the graphene. This lead to the formation of N-type semiconductors and opening in a small bandgap. However, the bandgap opening occurs due to symmetry breaking in the sublattices, hybridization, or a combination of both. The concentration of Alkaline atoms per unit-cell can tune the Fermi level shifting and bandgap. However, there is an optimal doping concentration to optimize the bandgap opening. In this study, we consider only one Alkaline atom per unit-cell; due to the interaction with one Alkaline atom, no additional bands appear near the Fermi level as decouple bilayer graphene band-structure.

### 3.2.1 Alkaline Metal Intercalation in Twisted Bilayer Graphene

This section shows the effects of Alkaline metal intercalation in the unit-cell of a twisted bilayer graphene, $C_8MC_8$. The significant bandgap opening and Fermi level shifting at K-point are highlighted in twisted $C_8MC_8$. Here, considered $1^0$, $2^0$, $3^0$, $4^0$, and $5^0$ twisted angles between carbon atoms with respect to the middle position of the $C_8MC_8$ structure as appeared in figure 1-(d). Figure 3-(b) represents how the bandgap varies to twist angles in the meV range. Therefore, this is the most useful graphene structure of FET and Infrared detector applications.

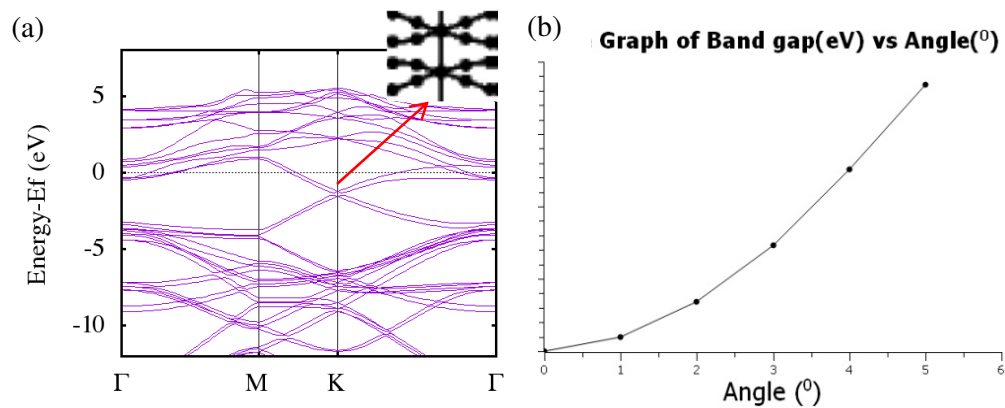

**Figure 3:** (a) The band diagram of $C_8CaC_8$ at 1° twist angle (b) The plot of band gap vs twist angle

### 3.3 Transport Properties of Twisted $C_8CaC_8$

The Seebeck coefficient depends on the doping level in semiconductors. The variation in the Seebeck coefficient with respect to $E - E_f$ for Ca intercalated graphene bilayer at 300 K is shown in Figure 4-(a). The Seebeck increases exponentially to the maximum, then decreases and vanishes around the bandgap with respect to doping level (Energy level) for all twisted angles, as shown in figure 4-(a). The Seebeck shows zero at the gap and the empty states. The lowest peak of Seebeck coefficient occurs for metal intercalated graphene bilayer without twisting, and the peak is increasing while increasing the twist angle between bilayers. The sign of the Seebeck coefficient changes at the charge neutrality point, which is charge density changes from electron to holes. The behavior of hall coefficient is the same as the Seebeck coefficient in the $C_8MC_8$ structure, as shown in figure 4-(b).

Figure 4–(c) & (d) represent the behavior of thermal conductivity and electrical conductivity vs $E - E_f$ for different twist angles of $C_8CaC_8$. Both the conductivity parameters decrease and go to zero until they reach its bandgap, and then increase. However, both conductivity is de*creased* when the twist angle increases due to an increase in the bandgap. The values are overestimated due to computational approximations and not considering phonon interaction in the calculation. However, these behaviors are promising to have thermoelectric materials in real applications.

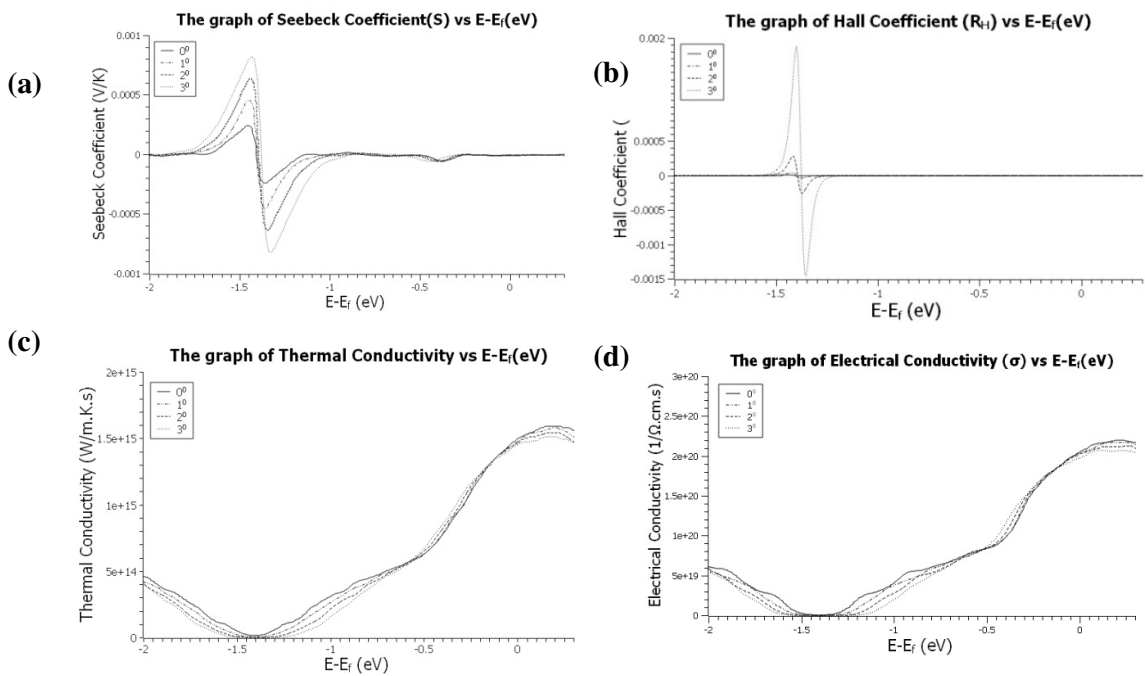

**Figure 4:** (a) Seebeck coefficient vs $E - E_f$  (b) Hall coefficient vs $E - E_f$ (c) Thermal conductivity vs $E - E_f$ (d) Electrical conductivity vs $E - E_f$ for twisted $C_8CaC_8$

## 4. CONCLUSIONS

Alkaline metal intercalated twisted bilayer graphene is a promising material for real-world application. According to the study, intercalation of metal ions to GL graphene, increases its bandgap and reduces the electrical conductivity, and thermal conductivity. The twisting of graphene bilayers increases in the Seebeck coefficient, Hall coefficient and reduces in thermal conductivity, & electrical conductivity for different twist angles. Further studies are required to find the optimized system and twist angles to increase the performance of the materials. This fundamental study focuses on understanding the properties of unit-cell, which can be expanded into supercell calculation, implementing phonon interaction as future works.

## 5. ACKNOWLEDGEMENT

We want to take this opportunity to thank Prof. Hai-Qing Lin (HQL), Director, Beijing Computational Science Research Center (CSRC). We are very grateful to thank Prof. HQL for facilitating supercomputer access at CSRC.